\documentclass[manuscript,screen]{acmart}

\AtBeginDocument{%
  \providecommand\BibTeX{{%
    \normalfont B\kern-0.5em{\scshape i\kern-0.25em b}\kern-0.8em\TeX}}}

\setcopyright{none}

\acmJournal{TOMS}

\usepackage{listings}
\usepackage{xcolor}
\usepackage{graphicx}
\usepackage{multirow}

\lstdefinelanguage{Julia}{
  morekeywords={abstract,break,case,catch,const,continue,do,else,elseif,end,export,false,for,function,immutable,import,importall,if,in,macro,module,quote,return,switch,true,try,type,typealias,using,while},
  sensitive=true,
  morecomment=[l]{\#},
  morecomment=[n]{\#=}{=\#},
  morestring=[b]",
  morestring=[b]',
  morestring=[b]"""
}

\begin{document}

\title[IntegrateUnitary.jl]{IntegrateUnitary.jl: A \texttt{Julia} package for symbolic integration over Haar measures}

\author{{\L}ukasz Pawela}
\email{lpawela@iitis.pl}
\affiliation{%
  \institution{Institute of Theoretical and Applied Informatics, Polish Academy of Sciences}
  \streetaddress{Ba{\l}tycka 5}
  \city{Gliwice}
  \postcode{44-100}
  \country{Poland}
}

\author{Zbigniew Pucha\l a}
\email{z.puchala@iitis.pl}
\affiliation{%
  \institution{Institute of Theoretical and Applied Informatics, Polish Academy of Sciences}
  \streetaddress{Ba{\l}tycka 5}
  \city{Gliwice}
  \postcode{44-100}
  \country{Poland}
}

\renewcommand{\shortauthors}{Pawela and Pucha\l a}

\begin{abstract}
Symbolic integration over the Haar measure of compact groups is a computational
cornerstone in quantum information science and random matrix theory. We present
\texttt{IntegrateUnitary.jl}, a comprehensive Julia package for computing exact expectations
of polynomial functions over a wide range of compact groups ($U(d)$, $O(d)$,
$Sp(d)$, and $SU(d)$ for balanced polynomials), circular and Gaussian ensembles,
Ginibre ensembles, permutation groups, random pure states, and unitary
$t$-designs. The package provides a fully open-source implementation of the
Weingarten calculus and Wick contractions with broad symbolic-$d$ support for
entry-wise and trace-polynomial integrals, while selected workflows currently
require concrete integer dimensions (including higher pure trace moments
$|\mathrm{tr}(U)|^{2k}$ for $k > 1$ and HCIZ with \texttt{SymbolicMatrix}
inputs, and direct matrix-valued integration of
\texttt{SymbolicMatrix}/\texttt{SymbolicMatrixProduct} expressions), automatic
asymptotic expansions, a high-level symbolic trace interface that reconstructs
Weingarten graphs from index-free expressions, and a bridge to
\texttt{ITensors.jl} for tensor network averaging. We discuss the underlying
algorithms, including the Murnaghan-Nakayama rule and symplectic-orthogonal
duality, and demonstrate that the package efficiently handles high-degree
moments and quantum information metrics.
\end{abstract}



\maketitle

\section{Introduction}
\label{sec:intro}
Integration over the Haar measure of compact groups is a cornerstone technique
in mathematical physics, with profound implications in random matrix theory
(RMT) \cite{mehta2004random, forrester2010log, livan2018introduction} and
quantum information theory (QIT) \cite{nielsen2010quantum, watrous2018theory}.
The ability to compute moments of random matrices enables the exact
characterization of ``typical'' properties of quantum systems, such as the
average entanglement entropy of subsystems, a concept famously pioneered by Page
\cite{page1993average}. Furthermore, these integrals are essential for studying
quantum chaos, scrambling, and the design of quantum circuits that approximate
randomness, known as unitary $t$-designs \cite{brandao2016exponential}.

Historically, the evaluation of these integrals was a formidable challenge.
While the defining properties of Haar measure were established by Hurwitz and
Weyl in the early 20$^\mathrm{th}$ century, a systematic method for evaluating
polynomial integrals emerged later with the work of Weingarten
\cite{weingarten1978asymptotic}. He showed that asymptotic integrals could be
expressed as sums over permutations involving a weight function now called the
Weingarten function. This theory was rigorously detailed and extended to finite
$d$ by Collins \cite{collins2003moments}. The seminal work of Collins and Śniady
\cite{collins2006integration} unified the treatment for the unitary, orthogonal,
and symplectic groups, providing explicit combinatorial formulas for the
Weingarten functions in terms of characters of the symmetric group (or related
structures).

Despite the existence of these closed-form solutions, their practical evaluation
remains non-trivial due to the combinatorial complexity of the symmetric group
sums and the intricacies of representation theory. Several software packages
have been developed to automate these calculations. Notable examples include
\texttt{RTNI} \cite{fukuda2019rtni}, a Mathematica package that offers robust
functionality but relies on the proprietary Wolfram engine, and \texttt{Haarpy}
\cite{cardin2024haarpy}, a Python library that provides accessible symbolic
integration. Earlier work by Puchała and Miszczak \cite{puchala2017symbolic}
also explored symbolic integration algorithms. A detailed feature comparison
is provided in Table~\ref{tab:comparison}, and a direct performance comparison
with \texttt{Haarpy} and \texttt{RTNI} is presented in
Sections~\ref{sec:haarpy_comparison} and~\ref{sec:rtni_comparison}.

However, existing tools often face limitations in extensibility, performance, or
the ability to handle symbolic dimensions $d$ natively across a broad class of workflows.
Furthermore, the translation from high-level, coordinate-free mathematical
expressions (e.g., traces of products) to the index-based tensor contractions
required by Weingarten calculus is often left to the user, leading to
error-prone manual preprocessing. \texttt{IntegrateUnitary.jl} addresses these issues and
focuses on maximizing performance and symbolic flexibility within the compact
group integration domain. A primary distinction lies in \texttt{IntegrateUnitary.jl}'s deep
integration with the Julia \texttt{Symbolics.jl} ecosystem, treating dimensions
$d$ as native symbolic variables rather than requiring separate
symbolic algebra backends or manual substitutions.
This diverse ensemble support, combined with specialized utilities for partial traces of symbolic subsystems, makes \texttt{IntegrateUnitary.jl} a versatile
environment for both theoretical derivations and complex numerical experiments.

\texttt{IntegrateUnitary.jl} offers a pure Julia implementation of the Weingarten calculus
for the unitary $U(d)$, orthogonal $O(d)$, and
symplectic $Sp(d)$ groups, as well as $SU(d)$ for balanced polynomials and
combinatorial integration for the symmetric group $S_d$. Key features distinguishing our approach include:
\begin{itemize}
    \item \textbf{Diverse ensemble support}: Unified integration over compact
    groups, circular ensembles, Gaussian ensembles (GUE, GOE, GSE), random pure
    states, diagonal unitary matrices (torus group), and unitary $t$-designs.
    \item \textbf{Symbolic dimensions and asymptotics}: Fully symbolic treatment of
    the dimension $d$, enabling the derivation of exact asymptotic expansions and
    universal scaling laws via automated Laurent series.
    \item \textbf{Symbolic trace logic}: A high-level interface that allows users to
    integrate trace polynomials directly, with the software automatically handling
    the reduction to Weingarten graphs.
    \item \textbf{Tensor network integration}: A native integration with
    \texttt{ITensors.jl} \cite{fishman2022itensor}, enabling the symbolic
    averaging of large-scale random tensor networks.
    \item \textbf{Partial trace}: A built-in function for computing partial
    traces of symbolic subsystems.
    \item \textbf{Performance}: Leveraging Julia's just-in-time compilation and
    efficient caching strategies to handle high-degree moments.
\end{itemize}

The remainder of this paper is organized as follows. Section
\ref{sec:background} reviews the mathematical background of Haar integration.
Section \ref{sec:features} describes the software architecture and core
features. Section \ref{sec:examples} provides usage examples, and Section
\ref{sec:implementation} details the implementation. We present benchmarks in
Section \ref{sec:performance} and conclude in Section \ref{sec:conclusion}.

\section{Background: Weingarten calculus}
\label{sec:background}
The core technique behind \texttt{IntegrateUnitary.jl} is the Weingarten calculus. For the
compact matrix groups considered here, the integral of a polynomial in
matrix elements can be expressed as a sum over a combinatorial set:
permutations for the unitary case and pair partitions for the orthogonal and
symplectic cases.

\subsection{The unitary group $U(d)$}
For the unitary group $U(d)$, the integral of a polynomial of degree $k$ in both
$U$ and $\bar{U}$ is given by:
\begin{equation}
\int_{U(d)} U_{i_1 j_1} \dots U_{i_k j_k} \bar{U}_{i'_1 j'_1} \dots \bar{U}_{i'_k j'_k} dU = \sum_{\sigma, \tau \in S_k} \delta_\sigma(\vec{i}, \vec{i}') \delta_\tau(\vec{j}, \vec{j}') \mathrm{Wg}^U(\sigma\tau^{-1}, d),
\end{equation}
where $S_k$ is the symmetric group of degree $k$, and $\delta_\sigma(\vec{i},
\vec{i}')$ is a product of Kronecker deltas $\prod_m \delta_{i_m,
i'_{\sigma(m)}}$. The Weingarten function $\mathrm{Wg}^U(\sigma, d)$ depends
only on the cycle type of $\sigma$ and, for generic $d$, can be expressed via
the irreducible characters $\chi_\lambda$ of $S_k$:
\begin{equation}
\mathrm{Wg}^U(\sigma, d) = \frac{1}{(k!)^2} \sum_{\lambda \vdash k} \frac{(f^\lambda)^2 \chi_\lambda(\sigma)}{s_\lambda(1^d)},
\end{equation}
where $f^\lambda$ is the dimension of the irreducible representation $\lambda$
and $s_\lambda(1^d) = s_\lambda(1,\ldots,1)$ is the Schur polynomial evaluated
at $d$ variables all equal to one, representing the dimension of the
corresponding $U(d)$ representation.

Crucially, $s_\lambda(1^d)$ is a polynomial in $d$, which implies that
$\mathrm{Wg}^U(\sigma, d)$ is a rational function of $d$. This property is
essential for symbolic integration, as it allows for exact results even when $d$
is treated as a variable. Possible poles occur at integers $d$ such that
$|d| < k$, reflecting the breakdown of the expansion for small dimensions.

For fixed $k$ and $\sigma$, in the asymptotic limit $d \to \infty$, the
Weingarten function satisfies:
\begin{equation}
\mathrm{Wg}^U(\sigma, d) = O(d^{-k - |\sigma|}),
\end{equation}
where $|\sigma|$ denotes the minimum number of transpositions required to
generate $\sigma$, related to the number of cycles $c(\sigma)$ by $|\sigma| =
k - c(\sigma)$. For the identity permutation, the leading order term is
$\mathrm{Wg}^U(\mathrm{id}, d) \approx d^{-k}$, recovering the normalization
expected from independent Gaussian entries at the naive limit.

\paragraph{Symbolic $d$ pitfalls}
While Weingarten functions are rational in $d$, they contain poles at small
integer dimensions (typically $d < k$ for degree $k$ moments). Furthermore,
substituting numeric values can yield $0/0$ expressions, which are removable
singularities, at specific points like $d=1, 2$. \texttt{IntegrateUnitary.jl}'s
\texttt{evaluate} function automatically simplifies these expressions to resolve
such singularities. An important exception is pure trace moments
$|\mathrm{tr}(U)|^{2k}$, whose exact value $\sum_{\lambda \vdash k,\,
\ell(\lambda) \le d} (f^\lambda)^2$ depends on $d$ as a step function rather
than a rational function; these require a concrete integer dimension and raise
an error for symbolic $d$.

\paragraph{Special Unitary group $SU(d)$}
For all currently supported ``balanced'' polynomial expressions (where the
number of $U$ and $\bar{U}$ factors are equal), the integration over $SU(d)$ is
equivalent to $U(d)$. Non-stable-range effects involving $\epsilon$-tensor
contractions for specific small $d$ are not currently covered.

\subsection{Orthogonal and symplectic groups}
For the orthogonal group $O(d)$ and compact symplectic group $Sp(d)$
(in the package convention, a $d \times d$ matrix group with even
$d = 2n$), odd moments vanish and even moments are indexed by the set of
pair partitions $P_{2k}$ of $\{1, \dots, 2k\}$. For $O(d)$, the integral is:
\begin{equation}
\int_{O(d)} O_{i_1 j_1} \dots O_{i_{2k} j_{2k}} dO = \sum_{p, q \in P_{2k}} \delta_p(\vec{i}) \delta_q(\vec{j}) \mathrm{Wg}^O(p, q, d).
\end{equation}
Here $\delta_p(\vec{i}) = \prod_{\{a,b\} \in p} \delta_{i_a i_b}$.
Similarly, for the symplectic group $Sp(d)$, with standard symplectic form
$J=\begin{pmatrix}0&I_n\\-I_n&0\end{pmatrix}$, the formula is:
\begin{equation}
\int_{Sp(d)} S_{i_1 j_1} \dots S_{i_{2k} j_{2k}} dS = \sum_{p, q \in P_{2k}} \Delta^J_p(\vec{i})\, \Delta^J_q(\vec{j})\, \mathrm{Wg}^{Sp}(p, q, d),
\end{equation}
where $\Delta^J_p(\vec{i}) = \prod_{(a,b) \in p,\ a < b} J_{i_a i_b}$
contracts each pair through the symplectic form $J$, introducing signs
$\pm 1$. For detailed closure relations on these groups, see Collins and
\'{S}niady
\cite{collins2006integration}.

\texttt{IntegrateUnitary.jl} leverages the fundamental duality between the orthogonal and
symplectic Weingarten functions:
    \begin{equation}
    \mathrm{Wg}^{Sp}(p, q, d) = (-1)^{\mathrm{loops}(p, q)} \mathrm{Wg}^O(p, q, -d).
    \end{equation}
where $\mathrm{loops}(p, q)$ is the number of loops in the graph formed by the
union of the two pair partitions $p$ and $q$. This relation allows the package
to unify the codebase, computing symplectic integrals by reusing the optimized
orthogonal engine with an appropriately signed dimension parameter $d \to -d$.

Both $\mathrm{Wg}^O$ and $\mathrm{Wg}^{Sp}$ are rational functions of $d$, and
their poles mark singularities of the rational inverse-Gram representation; the
Haar integrals themselves are well defined for admissible group dimensions.

\texttt{IntegrateUnitary.jl} efficiently computes these functions using character-based
formulas and Gram-matrix inversion (detailed in Section~\ref{sec:implementation}),
caching results to accelerate repeated integration tasks.

\subsection{Permutation group $S_d$}
We represent the symmetric group $S_d$ by $d \times d$ permutation matrices and
integrate with respect to the normalized counting measure. For a
monomial in the matrix entries $P_{i,j}$, repeated identical factors are first
collapsed (equivalently, we work with the set of distinct index pairs). The
integral is non-zero only if this deduplicated set is consistent with a
permutation (i.e., all row indices are distinct and all column indices are
distinct). The result is:
\begin{equation}
\int_{S_d} P_{i_1 j_1} \dots P_{i_k j_k} dP = \frac{(d-k)!}{d!},
\end{equation}
where $k$ is the number of distinct index pairs after the collapse.

\section{Software features}
\label{sec:features}
\texttt{IntegrateUnitary.jl} is built on top of the \texttt{Symbolics.jl} ecosystem
\cite{symbolicsjl}, providing a seamless experience for Julia users.

\subsection{Groups and measures}
The package supports a wide range of integration measures, grouped as follows:
\begin{itemize}
    \item \textbf{Compact groups}: Standard Haar measures on the unitary $U(d)$,
    special unitary $SU(d)$, orthogonal $O(d)$, and symplectic $Sp(d)$ groups.
    These are defined via \texttt{dU(d)}, \texttt{dSU(d)}, \texttt{dO(d)}, and
    \texttt{dSp(d)}, where the argument is the matrix dimension $d$; for
    concrete dimensions, \texttt{dSp(d)} requires even $d$. For balanced
    polynomials in the currently supported stable range, $SU(d)$ and $U(d)$
    integrals coincide; non-stable-range $\epsilon$-tensor effects at specific
    small $d$ are not currently covered.
    \item \textbf{Circular ensembles}: Measures for the circular unitary (CUE),
    orthogonal (COE), and symplectic (CSE) ensembles, accessed via
    \texttt{dCUE(d)}, \texttt{dCOE(d)}, and \texttt{dCSE(d)}. The constructor
    \texttt{dCUE(d)} is an alias for \texttt{dU(d)}, while \texttt{dCSE(d)}
    requires even concrete matrix dimensions; COE and CSE represent symmetric
    and self-dual unitary matrices, respectively.
    \item \textbf{Quantum states}: The Fubini--Study measure on pure states
    $|\psi\rangle$ drawn from the complex projective space $\mathbb{C}P^{d-1}$
    \cite{zyczkowski2001induced, bengtsson2017geometry}, accessed via
    \texttt{dPsi(d)}. This corresponds to the distribution of the first
    column of a Haar-random unitary matrix and is represented as a $(d, 1)$ symbolic matrix.
    \item \textbf{Permutation groups}: Measures for the Symmetric Group $S_d$
    (\texttt{dPerm(d)}) and the ensemble of centered permutation matrices
    (\texttt{dCPerm(d)}). Centered permutations $Y$ satisfy $Y_{ij} = P_{ij} - 1/d$, where $P \in S_d$.
    \item \textbf{Gaussian and Ginibre ensembles}: Support for the Gaussian
    unitary (GUE), orthogonal (GOE), and symplectic (GSE) ensembles, as well as
    the complex (GinUE), real (GinOE), and symplectic (GinSE) Ginibre ensembles,
    accessed via \texttt{dGUE(d)}, \texttt{dGOE(d)}, \texttt{dGSE(d)},
    \texttt{dGinUE(d)}, \texttt{dGinOE(d)}, and \texttt{dGinSE(d)}. For
    concrete dimensions, \texttt{dGSE(d)} and \texttt{dGinSE(d)} require even
    $d$. These rely on Wick's theorem~\cite{wick1950evaluation} for integration
    rather than Weingarten calculus.
    \item \textbf{Unitary designs}: Unitary $t$-designs (\texttt{dDesign(d, t)}),
    which mimic the first $t$ moments of the Haar measure, useful for studying
    pseudo-randomness in quantum circuits.
    \item \textbf{Stiefel manifolds}: The Stiefel manifold $V_k(\mathbb{C}^d)$,
    representing $d \times k$ matrices with orthonormal columns. This
    generalizes Haar-random pure states ($k=1$) and is implemented via
    \texttt{dStiefel(d, k)}, with $k \le d$ required for concrete dimensions.
    \item \textbf{Diagonal unitary matrices}: Integration over the torus group
    $T^d$ (\texttt{dDiagUnitary(d)}), representing independent phase averaging for
    each diagonal entry.
    \item \textbf{Matrix integration}: Native support for integrating
    AbstractArray-valued expressions with concrete integer result dimensions,
    enabling coordinate-free validation of matrix identities (e.g.,
    $\mathbb{E}[U U^\dagger] = I$).
\end{itemize}

\subsection{Harish-Chandra-Itzykson-Zuber (HCIZ) integrals}
Beyond polynomial moments, \texttt{IntegrateUnitary.jl} implements the closed-form
Harish-Chandra-Itzykson-Zuber (HCIZ) integral \cite{harish1958ranges,
itzykson1980planar}, a fundamental object in random matrix theory related to the
character expansion of the exponential function:
\begin{equation}
\int_{U(d)} dU e^{\text{Tr}(A U B U^\dagger)} = \left( \prod_{p=1}^{d-1} p! \right) \frac{\det(e^{a_i b_j})_{i,j=1}^d}{\Delta(a) \Delta(b)},
\end{equation}
where $a_i$ and $b_j$ are the eigenvalues of the source matrices $A$ and $B$,
and $\Delta(x)$ denotes the Vandermonde determinant. When the eigenvalues are
symbolic and non-degenerate, the evaluation is exact; for floating-point input
the result is numerical.

\texttt{IntegrateUnitary.jl} provides two primary interfaces for these integrals. The
\textbf{eigenvalue interface} allows users to pass vectors of eigenvalues
directly, facilitating purely symbolic derivations where the spectrum is defined
algebraically. The \textbf{matrix interface} accepts Hermitian matrices $A$ and
$B$ and dispatches on the element type. For numeric matrices, eigenvalues are
computed via standard diagonalization (\texttt{eigen}). For symbolic matrices
of type \texttt{Matrix\{Num\}}, the package extracts eigenvalues from diagonal
matrices of any dimension and from general $2 \times 2$ matrices via the
quadratic formula; larger non-diagonal symbolic matrices require the user to
supply eigenvalues directly. For \texttt{SymbolicMatrix} inputs with a concrete integer dimension,
where explicit entries are not available, the package introduces formal
eigenvalue symbols ($a_1, \ldots, a_d$, $b_1, \ldots, b_d$) and evaluates the
formula in terms of these symbols; symbolic dimensions are not supported here
because the formula requires constructing a finite set of eigenvalue symbols and
a $d \times d$ determinant. When degenerate eigenvalues are detected in numeric
input, the implementation first sorts both spectra (using a total order by real
then imaginary part) to make the procedure permutation-invariant, and then
applies independent perturbations
$a_i \to a_i + i\,\epsilon_a$ and $b_i \to b_i + i\,\epsilon_b$, where
$\epsilon_a = \max(\|a\|_\infty,\, 1)\cdot 10^{-12}$ and
$\epsilon_b = \max(\|b\|_\infty,\, 1)\cdot 10^{-12}$. This breaks exact
degeneracies in both Vandermonde denominators while introducing errors of order
$\mathcal{O}(\max(\epsilon_a,\epsilon_b))$ in the result, which is negligible for
double-precision arithmetic. For symbolic eigenvalues, degenerate cases must
be resolved analytically by the user (e.g., via L'H\^{o}pital's rule).

\subsection{Gaussian random matrix ensembles}
In addition to compact groups, \texttt{IntegrateUnitary.jl} supports integration over the
Gaussian ensembles (GUE, GOE, GSE). These are implemented using Wick's theorem
\cite{wick1950evaluation} (Isserlis' theorem) for moment contraction. For a
Gaussian Hermitian matrix $H$, the expectation values are determined by the
pairwise contractions:
\begin{itemize}
    \item \textbf{GUE}: $\langle H_{ij} \bar{H}_{kl} \rangle = \delta_{ik} \delta_{jl}$, leading to $\langle \mathrm{tr}(H^2) \rangle = d^2$, $\langle \mathrm{tr}(H^4) \rangle = 2d^3 + d$, and $\langle \mathrm{tr}(H^6) \rangle = 5d^4 + 10d^2$.
    \item \textbf{GOE}: $\langle H_{ij} H_{kl} \rangle = \delta_{ik} \delta_{jl} + \delta_{il} \delta_{jk}$ (for real symmetric $H$), leading to $\langle \mathrm{tr}(H^2) \rangle = d^2 + d$.
    \item \textbf{GSE}: For self-dual Hermitian matrices, the package utilizes
    the even-moment duality relation:
    \begin{equation}
    \langle \mathrm{tr}(H^k) \rangle_{\mathrm{GSE}}(d) = (-1)^{\frac{k}{2} + 1} \langle \mathrm{tr}(H^k) \rangle_{\mathrm{GOE}}(-d), \qquad k \in 2\mathbb{N},
    \end{equation}
    while odd moments vanish, $\langle \mathrm{tr}(H^{2m+1}) \rangle_{\mathrm{GSE}} = 0$.
    This implies $\langle \mathrm{tr}(H^2) \rangle = d^2 - d$.
\end{itemize}
This approach allows \texttt{IntegrateUnitary.jl} to compute moments for all three ensembles
while maintaining full support for symbolic dimensions $d$. For concrete integer
dimensions, the symplectic measure constructor enforces even size:
\texttt{dGSE(n)} raises an error when \texttt{n} is odd.

\subsection{Ginibre ensembles}
Moving beyond Hermitian matrices, \texttt{IntegrateUnitary.jl} supports the Ginibre
ensembles (GinUE, GinOE, GinSE) \cite{ginibre1965statistical}, which consist of
non-Hermitian matrices with independent and identically distributed (i.i.d.)
Gaussian entries. These are integrated using similar Wick contraction rules:
\begin{itemize}
    \item \textbf{GinUE}: $\langle G_{ij} \bar{G}_{kl} \rangle = \delta_{ik} \delta_{jl}$, leading to $\langle \mathrm{tr}(G G^\dagger) \rangle = d^2$, $\langle \mathrm{tr}(G G^\dagger)^2 \rangle = d^4 + d^2$, and $\langle \mathrm{tr}((G G^\dagger)^2) \rangle = 2d^3$. Only contractions between $G$ and its conjugate $\bar{G}$ are non-vanishing.
    \item \textbf{GinOE}: $\langle G_{ij} G_{kl} \rangle = \delta_{ik}
    \delta_{jl}$ for real entries.
    \item \textbf{GinSE}: Symplectic entries treated via duality relations. For
    concrete integer dimensions, \texttt{dGinSE(n)} requires even \texttt{n} and
    raises an error on odd \texttt{n}.
\end{itemize}
Like the beta ensembles, these rules are implemented to support both symbolic 
dimensions $d$ and high-level trace expressions.

\subsection{Circular ensembles}
\texttt{IntegrateUnitary.jl} also supports integration over the circular ensembles of random
unitary matrices, which are important in the study of quantum chaos and
symmetric spaces.
\begin{itemize}
    \item \textbf{CUE (Circular unitary ensemble)}: This corresponds to the
    standard Haar measure on $U(d)$. The matrices are unitary without further
    symmetry constraints.
    \item \textbf{COE (Circular orthogonal ensemble)}: Consists of symmetric
    unitary matrices ($S = S^T$). These are constructed as $S = U U^T$ where $U$
    is Haar-distributed on $U(d)$. \texttt{IntegrateUnitary.jl} handles integrals over COE
    by mapping the moments of $S$ to higher-order moments of $U$, reducing the
    problem to standard unitary Weingarten calculus.
    \item \textbf{CSE (Circular symplectic ensemble)}: Consists of self-dual
    unitary matrices ($S = U U^R$, where $U^R = J U^T J^T$ is the dual
    transpose). The integration is similarly performed by mapping to $U(d)$
    integrals (with $d$ even).
\end{itemize}
These are accessed via \texttt{dCOE(d)}, \texttt{dCSE(d)}, and
\texttt{dCUE(d)} (alias for \texttt{dU}).

\subsection{Unitary designs: Finite-moment measures and guards}
Unitary $t$-designs \cite{dankert2009exact, gross2007evenly,
ambainis2007quantum} are ensembles of unitary matrices that mimic the Haar
measure up to the $t^\mathrm{th}$ moment. Specifically, a set of unitaries
$\mathcal{D} \subset U(d)$ is a $t$-design if:
\begin{equation}
\mathbb{E}_{U \in \mathcal{D}} [P(U, \bar{U})] = \int_{U(d)} P(U, \bar{U}) dU
\end{equation}
for all polynomials $P$ of degree $q \le t$ in the entries of $U$ and $\bar{U}$.

\texttt{IntegrateUnitary.jl} provides the \texttt{dDesign(d, t)} measure to represent
such ensembles. The implementation enforces the moment-matching condition as
follows:
\begin{enumerate}
    \item For balanced integrands (equal degree $q$ in $U$ and $\bar{U}$) with
    $q \le t$, the package returns the exact Haar-averaged result using standard
    Weingarten calculus.
    \item For balanced integrands with $q > t$, the package raises an explicit
    error, preventing incorrect assumptions about the design's higher-moment
    behavior.
    \item Unbalanced integrands (different degree in $U$ and $\bar{U}$) return
    zero, which is the Haar-correct value. This is guaranteed correct for total
    degree $\le t$; for higher-degree unbalanced monomials, a $t$-design does
    not guarantee this value, but no error is currently raised.
\end{enumerate}
This feature allows researchers to verify whether specific quantum protocols or
randomized benchmarking schemes rely only on the $t$-moment properties of the
sampling distribution.

\subsection{Partial trace}
The package includes a \texttt{partial\_trace(M, dims, subsystem)} function
for quantum information tasks. It computes the partial trace of a matrix
$M$ over a specified subsystem, where the subsystem dimensions
(e.g., $d_A$, $d_B$ with $d = d_A d_B$) are concrete integers. The
matrix entries themselves may be symbolic expressions, for instance,
products of \texttt{SymbolicMatrix} elements, so the resulting reduced
density matrix retains symbolic dependencies suitable for further Haar
integration. This enables the calculation of entanglement metrics such as
purity by integrating the trace of the squared reduced density matrix, as
demonstrated in Section~\ref{sec:performance}.

\subsection{Symbolic dimensions and asymptotics}
A standout feature of \texttt{IntegrateUnitary.jl} is its deep support for symbolic dimensions $d$.
By leveraging the polynomial nature of group characters and Schur functions,
\texttt{IntegrateUnitary.jl} computes many supported element-wise and trace-of-product
integrals as exact rational functions of $d$; for supported
Haar/Weingarten-type matrix-group integrals, these take the form:
\begin{equation}
\int_{G(d)} P(U, \bar{U}) dU = \frac{N(d)}{D(d)},
\end{equation}
where $N(d)$ and $D(d)$ are polynomials. (Not all results are rational: pure trace moments $|\mathrm{tr}(U)|^{2k}$ depend on $d$ as a step function and require a concrete integer dimension; see Section~\ref{sec:limitations}.) This capability is critical for studying:
\begin{itemize}
    \item \textbf{Transition points}: Identifying poles in $D(d)$ that mark
    singularities of the rational representation; at admissible integer
    dimensions, removable singularities may still have well-defined values.
    \item \textbf{Thermodynamic limits}: The \texttt{asymptotic(expr, measure, order)}
    function computes the series expansion of the result in powers of $1/d$.
    This automates the extraction of large-$d$ behavior for rational-in-$d$
    observables (for example, high-degree entry moments and trace-polynomial
    integrals), where exact expressions can be combinatorially heavy.
    Pure trace moments $|\mathrm{tr}(U)|^{2k}$ are outside this workflow:
    they should be evaluated with \texttt{integrate(..., dU(n))} at concrete
    integer $n$.
\end{itemize}
Even when the input dimension is numeric (e.g., $d=3$), for measures supporting
symbolic reconstruction, the \texttt{asymptotic} routine can introduce a dummy
symbolic variable to perform the expansion for these rational-in-$d$
observables, providing theoretical insights alongside numerical results.

\subsection{Symbolic trace logic}
To simplify the integration of high-rank tensor networks and complex trace
expressions, \texttt{IntegrateUnitary.jl} introduces a \textbf{symbolic trace logic} system.
This system abstracts away explicit tensor indices, allowing users to define
computations in terms of coordinate-free matrix objects:
\begin{itemize}
    \item \textbf{Symbolic Matrix}: The \texttt{SymbolicMatrix} type represents
    an operator tagged by its role, such as a Haar unitary $U$, its adjoint
    $U^\dagger$, or a constant matrix $M$.
    \item \textbf{Lazy Evaluation}: Products such as $A * B$ build
    \texttt{SymbolicMatrixProduct} objects, while trace operations
    ($\mathrm{tr}$ or \texttt{tr\_lazy}) generate \texttt{LazyTrace} objects. These objects maintain
    algebraic structures (products of traces of matrix strings) without expanding
    indices, e.g., representing $\mathrm{tr}(U A U^\dagger B)$ as a graph cycle rather
    than a sum over four indices.
    \item \textbf{Graph Reconstruction}: During integration, the package
    interprets these lazy structures as contraction graphs. It automatically
    reconstructs the connectivity required for the relevant permutation,
    pair-partition, or Wick contractions, effectively converting the
    ``index-free'' input into the precise tensor contractions required by the
    relevant Weingarten or Wick formulas.
\end{itemize}
This abstraction significantly reduces the potential for index-mismatch errors
and renders the code nearly isomorphic to the mathematical pen-and-paper
formulation of the problem.

\subsection{ITensors.jl integration}
\label{sec:itensors}
Modern quantum circuit analysis and holographic duality models often rely on
tensor network representations. \texttt{IntegrateUnitary.jl} provides a specialized
extension for \texttt{ITensors.jl} \cite{fishman2022itensor}, allowing users to
integrate entire networks of \texttt{ITensor} objects symbolically.

To integrate an ITensor network, the user marks the random unitaries using the
\texttt{ITensorUnitary} wrapper, which specifies the tensor's input and output
indices. This explicit marking prevents ambiguities that arise from carrying
index tags (e.g., ``Haar'') over to constant tensors. The integration routine
then analyzes the contraction topology of the network and expands it into a sum
of deterministically contracted ITensors, each weighted by the appropriate
Weingarten function. This graphical Weingarten engine is significant for its
automatic handling of large, sparse networks where manual index expansion would
be computationally prohibitive.

\subsection{Comparison with existing tools}
\label{sec:comparison}
Table~\ref{tab:comparison} provides a detailed feature comparison of
\texttt{IntegrateUnitary.jl} with the two most closely related packages: \texttt{RTNI}
\cite{fukuda2019rtni}, a Mathematica package for integrating Haar-random tensor
networks, and \texttt{Haarpy} \cite{cardin2024haarpy}, a Python library for
Weingarten calculus over classical compact groups. The comparison highlights
three key distinctions:
\begin{enumerate}
    \item \textbf{Ensemble breadth}: Among the compared tools,
    \texttt{IntegrateUnitary.jl} supports Gaussian, Ginibre, and circular ensembles
    alongside compact groups, as well as discrete structures such as
    permutation groups, unitary $t$-designs, Stiefel manifolds, and diagonal
    unitaries.
    \item \textbf{Symbolic abstractions}: Neither \texttt{RTNI} nor
    \texttt{Haarpy} provides asymptotic $1/d$ expansions or
    \texttt{IntegrateUnitary.jl}-style symbolic trace logic. \texttt{Haarpy} does not
    provide matrix-valued integration, while \texttt{IntegrateUnitary.jl} provides direct
    matrix-valued integration for concrete-size array outputs and
    \texttt{RTNI} exposes graph/tensor outputs that typically require
    additional trace or scalarization
    post-processing for like-for-like scalar comparisons. These features
    significantly reduce the manual effort required for common calculations.
    \item \textbf{Platform}: \texttt{RTNI} requires the proprietary Mathematica
    engine (though a Python port exists), while \texttt{Haarpy} and
    \texttt{IntegrateUnitary.jl} are fully open-source. \texttt{IntegrateUnitary.jl}'s Julia
    implementation offers JIT compilation and native type specialization,
    providing performance advantages for large-scale symbolic computations.
    Direct timing comparisons with \texttt{Haarpy} and \texttt{RTNI} are
    presented in Sections~\ref{sec:haarpy_comparison}
    and~\ref{sec:rtni_comparison}.
\end{enumerate}

\begin{table*}[t]
\centering
\caption{Feature comparison of symbolic Haar integration packages.
$\checkmark$ = fully supported, $\sim$\textsuperscript{a}/$\sim$\textsuperscript{b} = partial support with caveat \textsuperscript{a}/\textsuperscript{b},
-- = not supported.}
\label{tab:comparison}
\begin{tabular}{llccc}
\hline
Category & Feature & \texttt{IntegrateUnitary.jl} & \texttt{RTNI} & \texttt{Haarpy} \\
\hline
\multirow{4}{*}{Groups}
  & $U(d)$             & $\checkmark$ & $\checkmark$ & $\checkmark$ \\
  & $SU(d)$ (balanced)  & $\checkmark$ & --           & --           \\
  & $O(d)$             & $\checkmark$ & --           & $\checkmark$ \\
  & $Sp(d)$            & $\checkmark$ & --           & $\sim$\textsuperscript{a}       \\
\hline
\multirow{3}{*}{Ensembles}
  & Circular (COE/CSE) & $\checkmark$ & --           & $\sim$\textsuperscript{a}       \\
  & Gaussian (GUE/GOE/GSE) & $\checkmark$ & --       & --           \\
  & Ginibre (GinUE/GinOE/GinSE) & $\checkmark$ & --  & --           \\
\hline
\multirow{5}{*}{Discrete/other}
  & Permutation groups  & $\checkmark$ & --           & $\checkmark$ \\
  & Unitary $t$-designs & $\checkmark$ & --           & --           \\
  & Stiefel manifolds   & $\checkmark$ & --           & --           \\
  & Diagonal unitaries  & $\checkmark$ & --           & --           \\
  & Random pure states  & $\checkmark$ & --           & --           \\
\hline
\multirow{3}{*}{Symbolic}
  & Symbolic dimension $d$ & $\checkmark$ & $\checkmark$ & $\checkmark$ \\
  & Asymptotic $1/d$ expansions & $\checkmark$ & --   & --           \\
  & Symbolic trace logic & $\checkmark$ & --           & --           \\
\hline
\multirow{3}{*}{Interfaces}
  & Tensor network integration & $\checkmark$ & $\checkmark$ & --   \\
  & Matrix-valued integration  & $\checkmark$ & $\sim$\textsuperscript{b}       & --   \\
  & HCIZ integrals             & $\checkmark$ & --           & --   \\
\hline
\multirow{3}{*}{Platform}
  & Language            & Julia        & Mathematica/Python & Python \\
  & Open-source runtime & $\checkmark$ & --           & $\checkmark$ \\
  & License             & Apache 2.0   & GPL v3.0     & Apache 2.0   \\
\hline
\end{tabular}
\\[4pt]
{\footnotesize \textsuperscript{a}Haarpy: Weingarten functions available but full integration not yet implemented (Sp($d$), CSE).\\
\textsuperscript{b}RTNI: graph/tensor-network outputs are available, but direct scalar value comparison typically requires extra trace/scalarization post-processing.}
\end{table*}

\subsection{Limitations and scope}
\label{sec:limitations}
The current implementation has several known limitations:
\begin{itemize}
    \item \textbf{$SU(d)$ scope}: Integration over $SU(d)$ is currently
    supported only for \emph{balanced} polynomial expressions, i.e., those
    containing equal numbers of $U$ and $\bar{U}$ factors. For such
    expressions, $SU(d)$ and $U(d)$ integrals coincide. Non-balanced
    integrals, which require $\epsilon$-tensor contractions and exhibit
    dimension-dependent behavior specific to small $d$, are not yet supported.
    In the current backend, these non-balanced queries are evaluated through
    the $U(d)$ rule and therefore return zero.
    \item \textbf{HCIZ degeneracies}: The Harish-Chandra-Itzykson-Zuber
    integral formula involves Vandermonde determinants in the denominator,
    which vanish when eigenvalues coincide. For numeric inputs with degenerate
    spectra, the package applies a small perturbation (see Section
    \ref{sec:features}); for symbolic inputs, degenerate cases currently
    require manual handling via L'H\^{o}pital's rule or limiting procedures.
    \item \textbf{Polynomial degree}: Due to the factorial growth of the
    symmetric group ($|S_k| = k!$) and pair partitions ($|M_{2k}| =
    (2k{-}1)!!$), exact symbolic integration is practically limited to
    polynomial degrees $2k \lesssim 12$ on standard hardware.
    \item \textbf{Trace moments}: Pure trace moments
    $|\mathrm{tr}(U)|^{2k}$ depend on $d$ as a step function (not a
    polynomial), so they require a concrete integer dimension.  For
    integer $d$, the library returns the exact value; for symbolic $d$,
    an \texttt{ArgumentError} is raised.
    \item \textbf{Matrix-valued integration}: Direct matrix-valued integration
    of \texttt{SymbolicMatrix} and \texttt{SymbolicMatrixProduct} expressions
    requires concrete integer result dimensions. If the output size is symbolic,
    users should scalarize the expression (for example with
    \texttt{tr(\ldots)}) instead of requesting a full matrix-valued result.
\end{itemize}

\section{Usage examples}
\label{sec:examples}
We illustrate the capabilities of \texttt{IntegrateUnitary.jl} with several examples, ranging
from basic component-wise integration to high-level symbolic trace manipulations
and asymptotic expansions. These examples demonstrate how the package's design
principles facilitate practical research tasks.

\subsection{Basic integration}
Consider the partial moment $\int_{U(d)} |U_{11}|^2 dU$. This integral represents
the average magnitude squared of a single entry of a Haar-random unitary matrix.
In \texttt{IntegrateUnitary.jl}, this can be computed directly with \texttt{@integrate}:
the macro infers both the symbolic matrix symbol (\texttt{U}) and the symbolic
dimension (\texttt{d}) from the measure context \texttt{dU(d)}:

\begin{lstlisting}[language=Julia]
using IntegrateUnitary
# The @integrate macro infers U and d from the measure context
result = @integrate abs(U[1,1])^2 dU(d)
# Output: 1/d

# Complex trace moments require a concrete integer dimension
res_tr = @integrate abs(tr(U))^4 dU(10)
# Output: 2

# Mixed-index fourth moment
res_c = @integrate U[1, 1] * conj(U[1, 2]) * U[2, 2] * conj(U[2, 1]) dU(d)
# Output: -1/(d*(d^2 - 1))

# Special Unitary group (stable range): matches U(d) for balanced moments
@integrate abs(U[1, 1])^2 dSU(d)
# Output: 1/d
\end{lstlisting}

The result $1/d$ confirms the expected normalization: since the rows and columns
of a unitary matrix are unit vectors, the average squared magnitude of any entry
must be $1/d$.

\subsection{Circular ensembles}
For the circular ensembles, we can observe distinct statistical properties
compared to the standard Haar measure. Here, we compute the second moment of a
diagonal entry for the symmetric (COE) and self-dual (CSE) cases:

\begin{lstlisting}[language=Julia]
# Circular Orthogonal Ensemble (S is symmetric)
@integrate abs(S[1, 1])^2 dCOE(d)
# Output: 2/(d+1)

# Circular Symplectic Ensemble (S is self-dual)
@integrate abs(S[1, 1])^2 dCSE(d)
# Output: 1/(d-1)

# Orthogonal group O(d): real matrix entries
@integrate O[1, 1]^2 dO(d)
# Output: 1/d

@integrate O[1, 1]^4 dO(d)
# Output: 3 / (d*(d+2))

# Symplectic group Sp(d): mixed moments
@integrate abs(Sp[1,1])^2 * abs(Sp[1,2])^2 dSp(d)
# Output: 1 / (d + d^2)   (poles at d = 0, -1)
\end{lstlisting}

These results contrast with the Haar unitary value of $1/d$, reflecting the
constraints imposed by the respective symmetries ($S=S^T$ for COE, $S=S^R$ for CSE, $O^TO=I$ for orthogonal, and the symplectic form for $Sp(d)$).

\subsection{Matrix integration}
\texttt{IntegrateUnitary.jl} allows for the direct integration of matrix-valued expressions
when the result dimensions are concrete integers. This eliminates the need for
manual element-wise iteration or broadcasting, automatically performing the
integration over all array entries. For Haar-random unitary matrices, we verify
the relation $\int U U^\dagger dU = I$:
\begin{lstlisting}[language=Julia]
using IntegrateUnitary
# Integrate the matrix product U * U'
# IntegrateUnitary handles the element-wise operations internally
@integrate U * U' dU(2)
# Output: Identity Matrix I
\end{lstlisting}

\subsection{Symbolic trace logic}
For more complex expressions involving traces of random matrices, manual index
contraction is error-prone. The symbolic trace interface simplifies the
workflow by allowing users to define the calculation in coordinate-free notation.
Here, we compute $\int \mathrm{tr}(U A U^\dagger B) dU$, a standard integral
appearing in the study of quantum channels (specifically, the definition of the
depolarizing channel):

\begin{lstlisting}[language=Julia]
# Compute integral of tr(U A U' B)
# The @integrate macro automatically treats unknown symbols as constant matrices
@integrate tr(U * A * U' * B) dU(d)
# Output: tr(A)tr(B)/d
\end{lstlisting}

This result neatly recovers the identity $\mathbb{E}[U A U^\dagger] =
\mathrm{tr}(A) \mathbb{I} / d$.

\subsection{Asymptotic expansions}
Analyzing behavior in the large-$d$ limit is crucial for understanding
thermodynamic properties and concentration of measure. \texttt{IntegrateUnitary.jl}
facilitates this analysis by expanding exact rational results into power series.
Consider the fourth moment of a matrix entry, $|U_{11}|^4$:

\begin{lstlisting}[language=Julia]
using IntegrateUnitary, Symbolics
@variables d
# The shorthand factory functions can be used for explicit variable creation
U = symbolic_unitary(:U, d)
expr = abs(U[1,1])^4
# Exact result computed internally: 2/(d*(d+1))
asymp_res = asymptotic(expr, dU(d), 4)
# Output: 2/d^2 - 2/d^3 + 2/d^4
\end{lstlisting}

This facility is particularly valuable for extracting universal scaling laws
from combinatorially complex exact expressions. For higher trace-polynomial
observables, it directly exposes both leading behavior and finite-size
corrections in powers of $1/d$:

\begin{lstlisting}[language=Julia]
using IntegrateUnitary, Symbolics
@variables d
U = symbolic_unitary(:U, d)
A = SymbolicMatrix(:A); B = SymbolicMatrix(:B)
C = SymbolicMatrix(:C); D = SymbolicMatrix(:D)

asymp_tp = asymptotic(tr(U*A*U'*B*U*C*U'*D), dU(d), 3)
# Leading terms:
# (tr(A)tr(B*D)tr(C) + tr(A*C)tr(B)tr(D))/d^2
# - (tr(A)tr(B)tr(C)tr(D) + tr(A*C)tr(B*D))/d^3
\end{lstlisting}

In contrast, pure trace moments have exact finite-$d$ dependence given by a
step function in integer $d$:
\[
\mathbb{E}\!\left[|\mathrm{tr}(U)|^{2k}\right]
= \sum_{\lambda \vdash k,\ \ell(\lambda)\le d} (f^\lambda)^2.
\]
Hence, for fixed $k$, the value equals $k!$ exactly once $d \ge k$ (for example,
$\mathbb{E}[|\mathrm{tr}(U)|^6]=6$ for all integer $d \ge 3$), and there is no
$1/d$-type subleading correction series in that stabilized regime.

As a physically grounded application, consider Page's result on the average
entanglement of random bipartite states \cite{page1993average}. For a random
pure state on $\mathbb{C}^n \otimes \mathbb{C}^n$, the average purity of
either subsystem is $\mathbb{E}[\mathrm{tr}(\rho_A^2)] = 2n/(n^2+1)$.
The \texttt{asymptotic} function accepts any rational expression, so we can
immediately extract the large-$n$ behavior:

\begin{lstlisting}[language=Julia]
using IntegrateUnitary, Symbolics
@variables n
page_purity = 2n / (n^2 + 1)
asymptotic(page_purity, n, 5)
# Output: 2/n - 2/n^3 + 2/n^5
\end{lstlisting}

The leading term $2/n$ shows that the subsystem purity is of order $1/n$ as the
dimension grows. For comparison, a maximally mixed $n$-dimensional state has
purity exactly $1/n$, so Page's law indicates highly mixed (and thus nearly
maximally entangled) reduced states rather than equality with the maximally
mixed value. The subleading
corrections $-2/n^3 + \cdots$ quantify the finite-size deviations from
maximal entanglement. Such expansions automate the derivation of scaling laws
where manual simplification of the underlying Weingarten sums would be
prohibitive.

\subsection{Tensor network integration}
Finally, we demonstrate the integration of a tensor network using the
\texttt{ITensors.jl} interface. This approach is particularly useful for Haar
averaging of random quantum circuits and tensor networks without explicit index
manipulation. We define a balanced network consisting of a random unitary $U$,
its adjoint $U^\dagger$, and constant tensors $A,B$, then compute its Haar
average.

\begin{lstlisting}[language=Julia]
using IntegrateUnitary, ITensors
i, j, k, l = Index(2), Index(2), Index(2), Index(2)

# Define U and its adjoint U_dag
U = ITensorUnitary(out_indices=[i], in_indices=[j])
U_dag = ITensorUnitary(out_indices=[k], in_indices=[l], is_adj=true)

# Constant tensors A and B to form a trace-like network
A = randomITensor(j, k); B = randomITensor(l, i)

# Integrate the balanced network [U, A, U_dag, B] over U(2)
integrate([U, A, U_dag, B], dU(2))
# result is a scalar ITensor from Weingarten delta contractions
# (for d=2 the overall prefactor is 1/2)
\end{lstlisting}

This high-level approach allows symbolic integration to be expressed directly at
the network level, avoiding manual elementwise index expansion while letting the
underlying engine operate on the network topology.
 
\subsection{Permutation groups}
Symbolic integration over the Symmetric Group $S_d$ is particularly useful for
combinatorial problems and studying centered permutation ensembles.
\begin{lstlisting}[language=Julia]
# Average of a product of entries
@integrate P[1, 1] * P[2, 2] dPerm(d)
# Output: 1 / (d * (d - 1))

# Centered permutations Y = P - J/d
@integrate Y[1, 1]^2 dCPerm(d)
# Output: (d - 1) / d^2
\end{lstlisting}

The first result, $1/(d(d-1))$, is the probability that a uniformly random
permutation maps both 1 and 2 to themselves, reflecting the well-known
inclusion-exclusion counting of derangements. The second result shows
that centring the permutation matrix removes the $1/d$ mean, leaving a
variance that converges to $1/d$ for large $d$.

\subsection{Diagonal unitary matrices}
For the group of diagonal unitary matrices (the torus $T^d$), integration
reduces to independent phase averaging. This arises naturally in the
dephasing channel and in studies of quantum coherence.
\begin{lstlisting}[language=Julia]
# Average of |D_11|^2
@integrate abs(D[1, 1])^2 dDiagUnitary(d)
# Output: 1
\end{lstlisting}

The result $1$ reflects the fact that each diagonal entry $D_{ii} = e^{i\theta_i}$
has unit modulus, so $|D_{11}|^2 = 1$ deterministically.
 
\subsection{Stiefel manifolds}
Integration over the Stiefel manifold $V_k(\mathbb{C}^d)$ is supported,
generalizing pure states to higher-rank orthonormal frames. This is crucial for
variational quantum algorithms and tensor network optimization
\cite{edelman1998geometry}.

\begin{lstlisting}[language=Julia]
# E[|V_{1,1}|^2]
@integrate abs(V[1, 1])^2 dStiefel(d, 2)
# Output: 1 / d
\end{lstlisting}

The result $1/d$ coincides with the unitary case because a random element of
$V_2(\mathbb{C}^d)$ is simply the first two columns of a Haar-random $U(d)$
matrix, so entries share the same second-moment structure.

\subsection{Ginibre ensembles}
Integration over Ginibre ensembles is useful for studying non-Hermitian random
matrix properties. We compute the second moment $\langle \mathrm{tr}(G
G^\dagger) \rangle$:
\begin{lstlisting}[language=Julia]
# Compute <tr(G G')>
@integrate tr(G * G') dGinUE(d)
# Output: d^2
\end{lstlisting}

The result $d^2$ follows from the fact that each entry of a complex Ginibre
matrix is an independent standard complex Gaussian, so $\mathbb{E}[|G_{ij}|^2] = 1$
and $\mathbb{E}[\mathrm{tr}(GG^\dagger)] = \sum_{i,j} \mathbb{E}[|G_{ij}|^2] = d^2$.

\section{Implementation details}
\label{sec:implementation}
\begin{figure*}[t]
\centering
\includegraphics[width=0.85\textwidth]{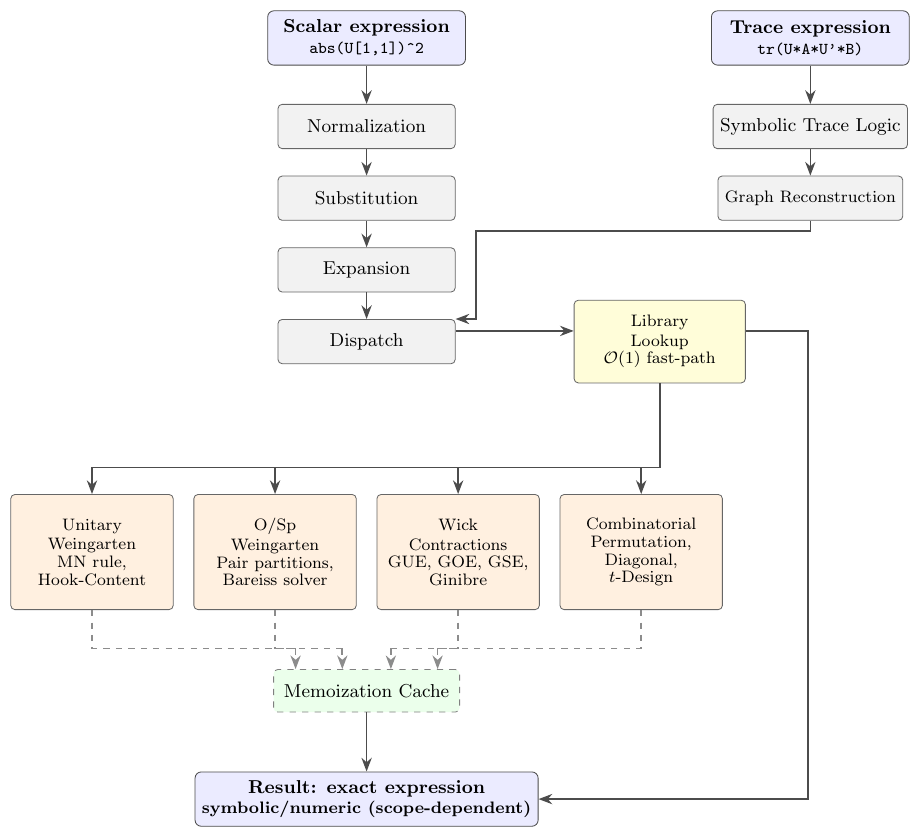}
\caption{Architecture of \texttt{IntegrateUnitary.jl}. Scalar expressions follow the
left path through normalization, substitution, expansion, and dispatch into
the library lookup. Trace expressions are first analyzed by the symbolic
trace logic, which reconstructs the Weingarten graph before entering the
same dispatch stage. Common patterns are returned directly by the library lookup
in $\mathcal{O}(1)$ time, while misses fall through to the specialized engines.
Performance-critical internals (representation-theoretic
quantities and Weingarten-related data) are memoized/cached for reuse.}
\Description{Architecture diagram of IntegrateUnitary.jl showing two flows: scalar expressions go through normalization, substitution, expansion, and dispatch into a library lookup; trace expressions first pass symbolic trace logic and graph reconstruction before the same dispatch stage. The library lookup returns common patterns in constant time, while remaining expressions fall through to specialized engines. Selected internal computations (for example characters, dimensions, and Weingarten data) are cached.}
\label{fig:architecture}
\end{figure*}

\texttt{IntegrateUnitary.jl} is written in pure Julia \cite{bezanson2017julia}, utilizing
the language's multiple dispatch and metaprogramming features to build a highly
modular and extensible system. The core architecture consists of three main
components: the integration pipeline, the Weingarten engine, and the symbolic
trace logic. Figure~\ref{fig:architecture} provides an overview of the data
flow.

\subsection{Integration pipeline}
The integration process follows a systematic pipeline designed to normalize and
dispatch symbolic expressions efficiently:
\begin{enumerate}
    \item \textbf{Normalization}: The input expression is first normalized using a ruleset
    based on \texttt{SymbolicUtils.jl}. This step expands complex conjugates and rewrites
    composite functions, such as $\mathrm{abs}(z)^2 \to z \bar{z}$, $\mathrm{real}(z) \to \frac{1}{2}(z +
    \bar{z})$, and $\mathrm{imag}(z) \to \frac{1}{2i}(z - \bar{z})$, into polynomial forms
    amenable to integration.
    \item \textbf{Substitution}: Symbolic variables representing matrix elements
    $U_{ij}$ are replaced by an internal atomic representation that facilitates
    pattern matching against the integration measure.
    \item \textbf{Expansion}: The expression is algebraically expanded into a sum of monomials.
    This relies on the efficient sparse polynomial handling of
    \texttt{Symbolics.jl}.
    \item \textbf{Dispatch}: Each monomial is analyzed to extract the indices of
    the random matrices. The integration is then dispatched to a centralized
    engine. This engine leverages the \texttt{AbstractMeasure} base type and a
    unified \texttt{measure\_info} interface, which provides the necessary
    metadata (substitution dictionaries, dimension parameters, and ensemble
    tags) for all supported measures. This architecture eliminates redundant
    boilerplate and ensures consistent dispatch behavior across the library.
\end{enumerate}

\subsection{Weingarten engine}
The calculation of Weingarten functions and combinatorial contractions is
centralized within a generic integration engine. To achieve high performance
and extensibility, \texttt{IntegrateUnitary.jl} implements specialized logic for different
mathematical structures while sharing the top-level dispatch mechanism:
\begin{itemize}
    \item \textbf{Unitary group}: Characters of the symmetric group are computed
    using the Murnaghan-Nakayama rule, which is significantly faster than determinantal
    formulas for sparse partitions. Dimensions of irreducible representations $s_\lambda(1^d)$
    are computed via the Hook-Content formula:
    \begin{equation}
    s_\lambda(1^d) = \prod_{(i,j) \in \lambda} \frac{d + c_{i,j}}{h_{i,j}}
    \end{equation}
    where $c_{i,j} = j-i$ is the content and $h_{i,j}$ is the hook length. This formulation
    allows the package to handle symbolic dimensions natively in supported workflows,
    producing exact rational functions of $d$ where applicable.
    \item \textbf{Orthogonal and symplectic groups}: For these groups, the package generates pair partitions
    recursively, canonicalizing them (typically as sorted pairs) to enable $\mathcal{O}(1)$ lookup.
    For symbolic dimensions $d$, \texttt{IntegrateUnitary.jl} utilizes a specialized univariate polynomial
    solver based on the Bareiss algorithm~\cite{bareiss1968sylvester} with exact \texttt{BigInt} arithmetic.
    Furthermore, to ensure stability for high-degree Weingarten sums where traditional
    simplification can fail, the engine employs a custom exact rational summation logic.
    Integration over the symplectic group reuses this optimized engine via duality:
    \begin{equation}
    \mathrm{Wg}^{Sp}(p, q, d) = (-1)^{\mathrm{loops}(p, q)} \mathrm{Wg}^O(p, q, -d).
    \end{equation}
    \item \textbf{Gaussian and Ginibre Ensembles}: Integration over Gaussian
    ensembles is performed via Wick contractions. The engine decomposes the
    monomial into atomic factors and generates all valid pair partitions of
    indices, weighted by the ensemble-specific contraction rules.
    \item \textbf{Diagonal unitary matrices}: For the torus group $T^d$, the
    integration is performed by verifying that the integrand is composed of
    diagonal elements and checking the multiset correspondence of indices
    between $D$ and $\bar{D}$ factors. This specialized engine avoids Weingarten
    functions entirely, providing significant performance gains for
    phase-averaging tasks.
    \item \textbf{Memoization}: All computed characters, dimensions, and
    Weingarten values are cached using the \texttt{Memoization.jl} package.
    Memoization occurs at the granular level of individual character and
    Weingarten function call, ensuring maximum reuse across different partition
    structures and integration orders.
\end{itemize}

\subsection{Symbolic trace logic}
To handle high-level expressions, \texttt{IntegrateUnitary.jl} introduces specialized
abstract types: \texttt{LazyTrace} and \texttt{LazySum}. These structures maintain
the algebraic form of trace products (e.g., $\mathrm{tr}(A B) \mathrm{tr}(C)$)
without strictly evaluating them into scalar components.

The integration strategy for these structures is twofold:
\begin{enumerate}
    \item \textbf{Library Lookup}: Common patterns, such as the single-channel
    overlap $\mathrm{tr}(U A U^\dagger B)$, are detected via pattern matching
    mechanisms (implemented in \texttt{check\_library}). These matches trigger optimized
    fast-paths that return pre-computed results directly (e.g.,
    $\frac{1}{d}\mathrm{tr}(A)\mathrm{tr}(B)$) without invoking the full Weingarten engine.
    This effectively reduces the complexity for these common cases from $O((k!)^2)$ to $O(1)$.
    \item \textbf{Graph Reconstruction}: For generic cases, the package
    interprets the trace cycles as edges in a graph. By identifying the relative
    positions of $U$ and $U^\dagger$ within the lazy structure, the integrator
    assigns effective indices to the intervening constant scaling matrices.
    This effectively converts the ``index-free'' input into the precise tensor
    contraction graph required by the Weingarten formula calculation, thereby
    avoiding the combinatorial explosion associated with eager index expansion.
\end{enumerate}

\paragraph{Input validation}
\texttt{IntegrateUnitary.jl} validates user inputs at the API boundary to provide clear
error messages. The package checks that Stiefel manifold parameters satisfy
$k \le d$, that partial trace subsystem indices are within range, and that
$t$-design integrals do not exceed the design order $t$. For the symplectic
group, dimensions must be even. For Haar-unitary rules (\texttt{dU}, and the
current \texttt{dSU} backend), unbalanced monomials are symmetry-forbidden and
evaluate to zero. \texttt{ArgumentError} is reserved for invalid measure
constraints (e.g., odd symplectic dimension, or a $t$-design query beyond the
design order).

\section{Performance and benchmarks}
\label{sec:performance}
We evaluated the performance of \texttt{IntegrateUnitary.jl} using a suite of stress tests
involving high-degree polynomial integrals over the unitary, orthogonal, and
symplectic groups. The benchmarks were run on a workstation equipped with an
Intel Core i7-12700KF processor (12 cores, 20 threads) and 64 GB of RAM (Julia
1.12.3). Specifically, we evaluated high-degree entry moments, such as
$\int |U_{11}|^{2k} dU$ for $2k \geq 8$, together with mixed-index and
tensor-network workloads. The entry-moment cases test high-degree symbolic
arithmetic and shortcut dispatch, while the mixed-index and tensor-network
cases exercise the general combinatorial contraction paths.

All native \texttt{IntegrateUnitary.jl} benchmarks in this section were performed using the
\texttt{BenchmarkTools.jl} package, and reported execution times represent the
median of $N=30$ cold-cache samples, with memoization caches cleared before
each sample. These results \emph{exclude} the initial JIT compilation and
warm-up time, reflecting steady-state performance observed during production
use. Cross-tool comparisons (Sections
\ref{sec:haarpy_comparison} and \ref{sec:rtni_comparison}) use dedicated
benchmark scripts with adaptive sample counts for very slow cases; those
policies are stated in the corresponding subsections. The reproduction script
\texttt{benchmarks/00\_manuscript\_benchmarks.jl} generates the native
\texttt{IntegrateUnitary.jl} benchmark tables in this section and can be executed via the
runner described in Section \ref{sec:availability}. Cross-tool comparisons
(Haarpy/RTNI) are generated by dedicated scripts in
\texttt{benchmarks/haarpy\_benchmarks/} and
\texttt{benchmarks/rtni\_benchmarks/}. Additional stress tests and correctness
checks are available in \texttt{benchmarks/09\_stress\_test\_2.jl} and
\texttt{benchmarks/08\_stress\_test\_1.jl}.

\subsection{Theoretical complexity}
The computational complexity of generic Weingarten integration is governed by
summation over the symmetric group $S_k$ (for the unitary case) or the set of
pair partitions $M_{2k}$ (for orthogonal/symplectic cases). The size of these
sets grows factorially: $|S_k| = k!$ and $|M_{2k}| = (2k-1)!!$. Consequently,
the uncompressed formulas scale exponentially with the degree of the
polynomial.

For the unitary group, the generic uncompressed formula contains a double sum
over $S_k$, giving $\mathcal{O}((k!)^2)$ terms per integral. The implementation
groups terms by cycle type where useful, and symmetric entry moments such as
$|U_{11}|^{2k}$ use closed-form or row/column moment shortcuts.
Individual Weingarten function evaluations require summing over partitions
$\lambda \vdash k$, each involving a character computation via the
Murnaghan-Nakayama rule in $\mathcal{O}(k^2)$ time per partition, and a
Hook-Content dimension evaluation in $\mathcal{O}(k)$ time. These values are
memoized, so repeated queries are $\mathcal{O}(1)$.

For the orthogonal and symplectic groups, the naive Gram matrix is indexed by
pair partitions and has $(2k-1)!! \times (2k-1)!!$ entries. The implementation's
main integration path instead solves a reduced system indexed by loop/cycle
types and sums grouped pair-partition contractions. The reduced Weingarten data
are cached once per degree and dimension, and subsequent queries reuse the
cached values. Symplectic integrals use the same reduced data with
$d \to -d$, together with the symplectic contraction signs, so they do not
require a separate full matrix inversion.

Empirically, we find that exact symbolic integration is feasible for degrees up
to $2k \approx 10\text{-}12$ on standard hardware. Beyond this regime, the number of
terms (e.g., $10! \approx 3.6 \times 10^6$) becomes prohibitive for real-time
symbolic manipulation, although numerical evaluation remains tractable for
slightly higher degrees.

\paragraph{Memory scaling}
Memory consumption is governed by two factors: the size of the intermediate
symbolic expressions and the memoization cache. For the unitary group, the
number of partitions $\lambda \vdash k$ grows sub-exponentially (the partition
function $p(k)$), and each cached Weingarten value is a single rational function
of $d$, so the cache remains modest (under 1~MiB for $k \le 6$). For the
orthogonal and symplectic groups, the uncompressed pair-partition matrix would
have $(2k{-}1)!! \times (2k{-}1)!!$ entries, but the main implementation stores
and reuses reduced loop/cycle-type data. The dominant memory cost of generic
intermediate symbolic expansions still grows rapidly with the number of valid
contractions. Once memoized, however, subsequent integrations at the same
degree $k$ incur little additional memory, as Weingarten data are retrieved from
cache rather than recomputed.

\subsection{Benchmark results}
Table \ref{tab:benchmarks} summarizes the execution times for selected integrals.
Key observations include:
\begin{enumerate}
    \item \textbf{Symbolic overhead}: Computing integrals with symbolic
    dimension $d$ is computationally demanding, but optimization strategies
    significantly reduce this cost. For example, computing $\int |U_{11}|^{10}
    dU$ symbolically took approximately 0.90 ms, achieving performance comparable
    to fixed $d=10$ numeric integration ($0.81$ ms). This efficiency leverages the
    grouping of identical cycle structures in the Weingarten sum.
    \item \textbf{Low-degree performance}: For common use cases involving
    lower-degree polynomials (e.g., $2k=6$), the package is extremely fast. The
    symbolic integration of $|U_{11}|^6$ completes in approximately 0.88 ms. For
    the orthogonal group, symbolic low-degree moments are similarly fast
    ($O_{11}^2$: 0.02 ms, $O_{11}^4$: 0.03 ms). For higher-degree orthogonal
    moments, the table reports concrete-dimension runs, where
    $O_{11}^{10}$ remains sub-millisecond at $d=20$ and $d=50$
    (0.37 ms and 0.36 ms).
    \item \textbf{Group differences}: Performance is workload-dependent.
    Unitary symbolic entry moments remain near the 1 ms range in this table,
    and orthogonal moments are highly efficient (especially in concrete
    high-degree cases). In contrast, the listed symplectic entry moments are
    noticeably slower at higher degree (4.51 ms for $|Sp_{11}|^8$ at $d=10$,
    and $\sim 78$--79 ms for $|Sp_{11}|^{10}$ at $d=10,20$). The dedicated
    comparison tables in Sections~\ref{sec:haarpy_comparison} and
    \ref{sec:rtni_comparison} show that the largest cross-tool speedups occur
    for high-degree unitary and orthogonal workloads, while some low-degree
    mixed-index unitary cases are near parity.
    \item \textbf{Correctness}: All benchmarks were verified against known exact
    values or asymptotic results, providing correctness checks for the reported
    cases.
    \item \textbf{Application}: The integration of the trace of the squared reduced density matrix for a bipartite state
    ($d=6$), which involves contracting a 4$^\mathrm{th}$ degree polynomial over
    unitary group, completes in approximately 9.81 ms, demonstrating the
    tool's applicability to quantum information tasks.
    \item \textbf{Tensor network scaling}: The graphical engine scales
    efficiently with both the degree $k$ and dimension $d$. As shown in Table
    \ref{tab:itensor_benchmarks}, for a loop network of Haar unitaries, the
    median time increases from 0.02 ms at $k=1$ to 15.44 ms at $k=4$ for
    fixed $d=2$. Dimension scaling for $k=2$ remains sub-second up to $d=50$,
    reaching 1.08 s at $d=100$. For higher degrees like $k=3$, the overhead
    increases more rapidly with $d$, with $d=30$ taking 17.05 s, reflecting the
    increasing rank of intermediate tensor contractions.
    \item \textbf{Permutation groups}: Integration over the symmetric group
    $S_d$ is exceptionally fast due to its combinatorial nature. Computing the
    degree-10 monomial $\prod_{i=1}^{10} P_{ii}$ (ten distinct index pairs) for
    $d=100$ takes only 0.49 ms. This
    high performance extends to the centered permutation ensemble, which remains
    efficient even after polynomial expansion.
\end{enumerate}

\begin{table}[ht]
\centering
\caption{Benchmark results for symbolic vs.\ numeric integration. Times are
median values over 30 samples.}
\label{tab:benchmarks}
\begin{tabular}{llcc}
\hline
Group & Integrand & Dimension $d$ & Time (ms) \\
\hline
Unitary & $|U_{11}|^6$ & Symbolic & 0.88 \\
Unitary & $|U_{11}|^8$ & Symbolic & 0.90 \\
Unitary & $|U_{11}|^{10}$ & Symbolic & 0.90 \\
Unitary & $|U_{11}|^{10}$ & $d=10$ & 0.81 \\
Unitary & $|U_{11}|^{10}$ & $d=50$ & 0.83 \\
\hline
Orthogonal & $O_{11}^2$ & Symbolic & 0.02 \\
Orthogonal & $O_{11}^4$ & Symbolic & 0.03 \\
Orthogonal & $O_{11}^6$ & $d=10$ & 0.37 \\
Orthogonal & $O_{11}^8$ & $d=20$ & 0.38 \\
Orthogonal & $O_{11}^{10}$ & $d=20$ & 0.37 \\
Orthogonal & $O_{11}^{10}$ & $d=50$ & 0.36 \\
\hline
Symplectic & $|Sp_{11}|^8$ & $d=10$ & 4.51 \\
Symplectic & $|Sp_{11}|^{10}$ & $d=10$ & 77.53 \\
Symplectic & $|Sp_{11}|^{10}$ & $d=20$ & 79.18 \\
\hline
GinUE & $\mathrm{tr}(GG^\dagger)$ & Symbolic & 0.00 \\
GinUE & $\mathrm{tr}(GG^\dagger)$ & $d=4$ & 0.00 \\
\hline
Circ. Orthogonal & $|S_{11}|^2$ & Symbolic & 0.02 \\
Circ. Orthogonal & $|S_{11}|^4$ & Symbolic & 0.03 \\
Circ. Orthogonal & $|S_{11}|^6$ & Symbolic & 1.07 \\
Circ. Symplectic & $|S_{11}|^2$ & Symbolic & 0.30 \\
Circ. Symplectic & $|S_{11}|^4$ & Symbolic & 2.92 \\
Circ. Symplectic & $|S_{11}|^6$ & Symbolic & 102.43 \\
\hline
Permutation & $\prod_{i=1}^{10} P_{ii}$ & $d=100$ & 0.49 \\
Permutation & $\mathrm{tr}(PA)^2$ & $d=4$ & 14.57 \\
Centered Perm. & $Y_{11}^4$ & $d=10$ & 0.43 \\
\hline
Application & Bipartite & $d=6$ & 9.81 \\
\hline
\end{tabular}
\end{table}

The results demonstrate that \texttt{IntegrateUnitary.jl} remains efficient for practically
relevant problem sizes. The rapid execution of the bipartite state application ($\approx$
9.81 ms) highlights the library's utility for routine quantum information tasks,
while the sub-millisecond performance for low-degree orthogonal integrals
facilitates large-scale statistical sampling. We further note that circular
ensemble timings are heterogeneous: COE rows remain near 0.02--1.07 ms in
Table~\ref{tab:benchmarks}, while higher-degree CSE examples are
substantially more expensive. For highly symmetric
integrands, such as powers of a single matrix entry, the integration engine
utilizes an optimized grouped-summation technique that significantly reduces
symbolic overhead by collecting terms with identical cycle types. Consequently,
high-degree moments such as $|O_{11}|^6$ are computed in under 0.40 ms.
Simultaneously, the symbolic engine proves capable of handling high-degree
moments that are intractable by hand. The caching mechanism ensures that
subsequent calls with the same parameters are virtually instantaneous.

\begin{table}[ht]
\centering
\caption{Benchmark results for ITensor network integration (Haar measure)
showing scaling with degree $k$ and dimension $d$. Timings exclude construction
of the random ITensor constants and index objects.}
\label{tab:itensor_benchmarks}
\begin{tabular}{lccc}
\hline
Scaling type & Degree $k$ & Dimension $d$ & Time (ms) \\
\hline
Degree $k$ (U) & 1 & 2 & 0.01 \\
Degree $k$ (U) & 2 & 2 & 0.05 \\
Degree $k$ (U) & 3 & 2 & 0.42 \\
Degree $k$ (U) & 4 & 2 & 13.02 \\
\hline
Degree $k$ (O) & 2 & 3 & 0.01 \\
Degree $k$ (O) & 4 & 3 & 0.33 \\
Degree $k$ (O) & 6 & 3 & 463.25 \\
\hline
Dimension $d$ ($k=2$) & 2 & 2 & 0.05 \\
Dimension $d$ ($k=2$) & 2 & 10 & 0.22 \\
Dimension $d$ ($k=2$) & 2 & 50 & 73.42 \\
Dimension $d$ ($k=2$) & 2 & 100 & 1009.99 \\
\hline
Dimension $d$ ($k=3$) & 3 & 2 & 0.41 \\
Dimension $d$ ($k=3$) & 3 & 10 & 14.49 \\
Dimension $d$ ($k=3$) & 3 & 20 & 1470.44 \\
Dimension $d$ ($k=3$) & 3 & 30 & 16141.72 \\
\hline
Orthogonal & 6 & 3 & 416.86 \\
\hline
\end{tabular}
\end{table}

\subsection{Matrix integration benchmarks}
We evaluated the performance of generic matrix integration by computing
$\mathbb{E}[U U^\dagger]$ for $N \times N$ symbolic matrices over $U(d)$. The
results (Table \ref{tab:matrix_bench}) demonstrate efficient scaling, as IntegrateUnitary
automatically handles the element-wise operations.

\begin{table}[h]
\centering
\caption{Benchmark results for matrix integration $\mathbb{E}[U U^\dagger]$ over
$U(d)$. Timings are for constructing and integrating the full $N \times N$
matrix of expressions.}
\label{tab:matrix_bench}
\begin{tabular}{ccc}
\hline
Matrix size ($N$) & Median time (ms) & Allocations (MiB) \\
\hline
$2 \times 2$ & 2.55 & 0.82 \\
$3 \times 3$ & 6.58 & 2.22 \\
$4 \times 4$ & 12.53 & 4.84 \\
\hline
\end{tabular}
\end{table}

\subsection{Performance comparison with Haarpy}
\label{sec:haarpy_comparison}
To provide a direct quantitative comparison with existing tools, we benchmarked
\texttt{IntegrateUnitary.jl} against \texttt{Haarpy} \cite{cardin2024haarpy}, the most
closely comparable open-source package. Both packages were evaluated on the same
machine using identical integrands. We tested both \emph{diagonal} moments
($|M_{11}|^{2k}$), which admit closed-form shortcuts, and \emph{off-diagonal}
products (e.g., $|U_{11}|^2|U_{12}|^2$), which exercise the general Weingarten
summation path. On the \texttt{IntegrateUnitary.jl} side, each row reports the median over
fixed $N=30$ cold-cache samples, with memoization caches cleared before every
sample. On the \texttt{Haarpy} side, each row reports the median with default
$N=30$ samples. The script includes probe-based adaptive reduction to $N=5$
for very slow cases, but this rerun stayed at $N=30$ for all reported rows.
Timings exclude JIT warm-up for \texttt{IntegrateUnitary.jl} and import overhead for
\texttt{Haarpy}. Speedups are computed from these per-row medians. The
comparison scripts are available in
\texttt{benchmarks/haarpy\_benchmarks/}.

Table~\ref{tab:haarpy_comparison} summarizes the results. For unitary integrals,
\texttt{IntegrateUnitary.jl} achieves strong speedups for diagonal moments ($|U_{11}|^{2k}$,
$k = 3, 4, 5$): $12.0\times$, $27.5\times$, and $88.5\times$ in the symbolic-$d$
rows, and $47.5\times$-$55.3\times$ in the numeric-$d$ rows. Off-diagonal
unitary integrals range from near parity ($1.5\times$ for
$|U_{11}|^2|U_{22}|^2$) to substantial gains ($11.4\times$ for
$|U_{11}|^4|U_{12}|^4$).

The orthogonal group results reveal a more dramatic separation: while
both packages handle low-degree cases efficiently, \texttt{IntegrateUnitary.jl} is already
substantially faster ($O_{11}^2$: $34.7\times$, $O_{11}^4$: $70.0\times$),
\texttt{IntegrateUnitary.jl}'s optimized univariate Weingarten solver delivers speedups
exceeding $2.7\times 10^4\times$ at degree 10. This disparity arises because
\texttt{Haarpy} constructs and inverts the full $(2k{-}1)!! \times (2k{-}1)!!$
Weingarten matrix even when a single entry suffices, whereas \texttt{IntegrateUnitary.jl}
exploits the structure of univariate integrands to bypass this cost entirely.
Off-diagonal orthogonal integrals show a $4.5$-$6.0\times$ advantage.

For the circular orthogonal ensemble (COE), \texttt{IntegrateUnitary.jl} uses a
closed-form expression for diagonal moments, yielding $20.9$-$127.2\times$
speedups that grow with the degree. We note that \texttt{Haarpy} (v0.0.6, the latest release available at
time of writing) returned incorrect results (zero) for the off-diagonal COE
integrals tested ($|S_{12}|^{2k}$, $|S_{11}|^2|S_{12}|^2$), which
\texttt{IntegrateUnitary.jl} evaluates correctly; these cases are therefore excluded
from the table.

Overall, \texttt{IntegrateUnitary.jl} demonstrates a substantial performance
advantage across most tested groups, with the largest gains in high-order
orthogonal and unitary diagonal moments.

\begin{table}[ht]
  \centering
  \caption{Performance comparison: \texttt{IntegrateUnitary.jl} vs.\ \texttt{Haarpy}
  (IntegrateUnitary.jl: fixed $N=30$ cold-cache samples; Haarpy: default $N=30$ with
  probe-based reduction to $N=5$ enabled, not triggered in this run; speedup
  $=$ Haarpy / IntegrateUnitary.jl from row
  medians). Both packages were run on the same hardware; timings exclude
  JIT/import overhead.}
  \label{tab:haarpy_comparison}
  \begin{tabular}{llrrr}
    \hline
    Group & Integrand & IntegrateUnitary.jl (ms) & Haarpy (ms) & Speedup \\
    \hline
    \multicolumn{5}{l}{\emph{Diagonal moments}} \\
    U($d$) & $|U_{11}|^6$, symbolic $d$ & 0.99 & 11.85 & 12.0$\times$ \\
    U($d$) & $|U_{11}|^8$, symbolic $d$ & 0.88 & 24.10 & 27.5$\times$ \\
    U($d$) & $|U_{11}|^{10}$, symbolic $d$ & 0.89 & 78.65 & 88.5$\times$ \\
    U($d$) & $|U_{11}|^{10}$, $d=10$ & 1.00 & 47.38 & 47.5$\times$ \\
    U($d$) & $|U_{11}|^{10}$, $d=50$ & 0.82 & 45.29 & 55.3$\times$ \\
    \hline
    O($d$) & $O_{11}^2$, symbolic $d$ & 0.02 & 0.74 & 34.7$\times$ \\
    O($d$) & $O_{11}^4$, symbolic $d$ & 0.06 & 3.89 & 70.0$\times$ \\
    O($d$) & $O_{11}^6$, $d=10$ & 0.36 & 0.91 & 2.5$\times$ \\
    O($d$) & $O_{11}^8$, $d=20$ & 0.37 & 63.17 & 172.7$\times$ \\
    O($d$) & $O_{11}^{10}$, $d=20$ & 0.35 & 9851.39 & 27942.2$\times$ \\
    O($d$) & $O_{11}^{10}$, $d=50$ & 0.53 & 9756.51 & 18408.3$\times$ \\
    \hline
    COE & $|S_{11}|^2$, symbolic $d$ & 0.02 & 2.37 & 107.7$\times$ \\
    COE & $|S_{11}|^4$, symbolic $d$ & 0.04 & 4.97 & 127.2$\times$ \\
    COE & $|S_{11}|^6$, symbolic $d$ & 0.95 & 19.85 & 20.9$\times$ \\
    \hline
    \multicolumn{5}{l}{\emph{Off-diagonal products}} \\
    U($d$) & $|U_{11}|^2|U_{12}|^2$, symbolic $d$ & 1.50 & 6.42 & 4.3$\times$ \\
    U($d$) & $|U_{11}|^4|U_{12}|^4$, symbolic $d$ & 1.90 & 21.71 & 11.4$\times$ \\
    U($d$) & $|U_{11}|^2|U_{22}|^2$, symbolic $d$ & 2.06 & 3.10 & 1.5$\times$ \\
    \hline
    O($d$) & $O_{11}^2 O_{12}^2$, symbolic $d$ & 0.60 & 3.61 & 6.0$\times$ \\
    O($d$) & $O_{11}O_{12}O_{21}O_{22}$, sym.\ $d$ & 1.09 & 4.90 & 4.5$\times$ \\
    \hline
\end{tabular}
\end{table}

\subsection{Performance comparison with RTNI}
\label{sec:rtni_comparison}
We benchmark \texttt{IntegrateUnitary.jl} against \texttt{RTNI} \cite{fukuda2019rtni} on
identical integrands over $U(d)$, executed on the same hardware.
Table~\ref{tab:rtni_comparison} reports median runtimes (default $N=10$
samples, adaptively reduced to $N=3$ for slow cases; JIT/import overhead
excluded). The speedup is defined as $\mathrm{RTNI} / \mathrm{IntegrateUnitary.jl}$, so
values greater than one indicate faster execution in \texttt{IntegrateUnitary.jl}.
All rows produce fully evaluated scalar results in both tools, ensuring an
apples-to-apples comparison. For \texttt{RTNI}'s graph-based engine, this
includes the cost of converting the internal graph representation to a scalar
expression via \texttt{converttomonomial}. Reproduction scripts are available
in \texttt{benchmarks/rtni\_benchmarks/}.

\begin{table*}[t]
  \centering
  \caption{Performance comparison: \texttt{IntegrateUnitary.jl} vs.\ \texttt{RTNI}
  (Mathematica). Median runtime over repeated runs. All rows produce
  fully evaluated scalar results in both tools.}
  \label{tab:rtni_comparison}
  \begin{tabular}{llrrr}
    \hline
    Group & Integrand & IntegrateUnitary.jl (ms) & RTNI (ms) & Speedup \\
    \hline
    U($d$) (Element API) & $|U_{11}|^2$, symbolic $d$ & 1.34 & 0.53 & 0.4$\times$ \\
    U($d$) (Element API) & $|U_{11}|^4$, symbolic $d$ & 1.49 & 2.51 & 1.7$\times$ \\
    U($d$) (Element API) & $|U_{11}|^6$, symbolic $d$ & 1.42 & 37.81 & 26.5$\times$ \\
    U($d$) (Element API) & $|U_{11}|^8$, symbolic $d$ & 1.41 & 5275.56 & 3745.2$\times$ \\
    U($d$) (Element API) & $|U_{11}|^8$, $d=10$ & 1.33 & 5123.40 & 3852.1$\times$ \\
    U($d$) (Element API) & $|U_{11}|^2|U_{12}|^2$, symbolic $d$ & 2.26 & 2.49 & 1.1$\times$ \\
    U($d$) (Element API) & $|U_{11}|^2|U_{12}|^4$, symbolic $d$ & 2.42 & 36.29 & 15.0$\times$ \\
    U($d$) (Element API) & $|U_{11}|^2|U_{22}|^2$, symbolic $d$ & 2.85 & 2.48 & 0.9$\times$ \\
    U($d$) (Element API) & $|U_{11}|^2|U_{12}|^2$, $d=10$ & 2.32 & 2.46 & 1.1$\times$ \\
    U($d$) (Element API) & $|U_{11}|^2|U_{12}|^4$, $d=10$ & 2.27 & 35.42 & 15.6$\times$ \\
    U($d$) (Element API) & $|U_{11}|^2|U_{22}|^2$, $d=10$ & 2.47 & 2.43 & 1.0$\times$ \\
    U($d$) & $|\mathrm{tr}(U)|^4$, $d=10$ & 0.01 & 0.18 & 19.9$\times$ \\
    U($d$) & $|\mathrm{tr}(U)|^6$, $d=10$ & 0.01 & 0.19 & 16.9$\times$ \\
    U($d$) & $|\mathrm{tr}(U)|^8$, $d=10$ & 0.02 & 0.21 & 12.0$\times$ \\
    U($d$) & $\mathrm{tr}(UAU^\ast B)$, symbolic $d$ & 0.13 & 0.50 & 3.9$\times$ \\
    U($d$) & $\mathrm{tr}((UAU^\ast B)^2)$, symbolic $d$ & 1.37 & 2.45 & 1.8$\times$ \\
    \hline
  \end{tabular}
\end{table*}

For element-wise integrals, the performance gap grows dramatically with
polynomial degree. At the lowest order ($|U_{11}|^2$), \texttt{RTNI} is
moderately faster ($0.3\times$). By degree eight ($|U_{11}|^8$),
\texttt{IntegrateUnitary.jl} is over three orders of magnitude faster
($3365\times$-$4072\times$), reflecting \texttt{IntegrateUnitary.jl}'s optimized
cycle-grouping strategy that avoids the full permutation-level expansion. For
mixed-index integrals, performance is near parity at low degree
($0.9\times$-$1.0\times$ for $|U_{11}|^2|U_{12}|^2$) and diverges at higher
degree ($14\times$-$15\times$ for $|U_{11}|^2|U_{12}|^4$). For trace-polynomial
integrals, $\mathrm{tr}(UAU^\ast B)$ and $\mathrm{tr}((UAU^\ast B)^2)$,
\texttt{IntegrateUnitary.jl} shows advantages of $2.8\times$ and $1.9\times$,
respectively.

\section{Conclusion}
\label{sec:conclusion}
We have presented \texttt{IntegrateUnitary.jl}, a robust and extensible Julia package for
performing symbolic integration over the Haar measure of the unitary,
orthogonal, and symplectic groups (and $SU(d)$ for balanced polynomials). By bridging the gap between the
abstract theory of Weingarten calculus and practical computation,
\texttt{IntegrateUnitary.jl} empowers researchers to rigorously derive properties of random
quantum channels, higher-order entanglement statistics, and scrambling measures
without resorting to brittle or approximate numerical sampling.

The package's core strengths lie in its unified treatment of compact groups,
its broad symbolic-$d$ support for entry-wise and trace-polynomial integrals
(with explicit concrete-$d$ exceptions for higher pure trace moments, HCIZ
on \texttt{SymbolicMatrix} inputs, and direct matrix-valued integration of
\texttt{SymbolicMatrix}/\texttt{SymbolicMatrixProduct} expressions), and its
novel symbolic trace logic. The latter effectively decouples the user's high-level
mathematical intent from the low-level index contractions required by the integration
engine, significantly reducing the cognitive load and potential for error. As
demonstrated by the benchmarks, the optimized Julia implementation ensures that
this symbolic abstraction does not come at the cost of performance, enabling the
evaluation of high-degree moments that were previously intractable.

Future development of \texttt{IntegrateUnitary.jl} will focus on expanding support to
exceptional Lie groups (particularly $G_2$, which arises in holonomy
classification and string-theoretic compactifications), integrating finite-$d$
corrections for $SU(d)$ beyond the stable range via $\epsilon$-tensor
contractions, and further optimizing the asymptotic expansion engine for
extremely large-scale problems. We believe \texttt{IntegrateUnitary.jl} will become an
essential utility in the toolkit of theoretical physicists and quantum
information scientists.

\section{Software availability and reproducibility}
\label{sec:availability}
The source code for \texttt{IntegrateUnitary.jl} is available on GitHub under the Apache License 2.0.
\begin{itemize}
    \item \textbf{Repository URL}: \url{https://github.com/iitis/IntegrateUnitary.jl}
    \item \textbf{Archived release DOI}: \url{https://doi.org/10.5281/zenodo.20346848}
    \item \textbf{License}: Apache License 2.0
    \item \textbf{Version}: The results in this paper were generated using
    version v1.0.0.
    \item \textbf{Pinned Julia environments}: Reproducibility relies on
    versioned \texttt{Manifest.toml} files in \texttt{examples/} and
    \texttt{benchmarks/}. The benchmark/example runner scripts instantiate
    these environments before execution.
    \item \textbf{Benchmarks}: Native IntegrateUnitary Julia benchmarks are turnkey from a
    clean checkout (with pinned Julia manifests). To reproduce them, run:
    \begin{lstlisting}[language=bash]
    bash benchmarks/runbenchmarks.sh
    \end{lstlisting}
    Cross-tool comparisons (\texttt{Haarpy}, \texttt{RTNI}) require manual
    external setup and are not one-command reproducible from Julia manifests
    alone.

    Haarpy setup and run:
    \begin{lstlisting}[language=bash]
    cd benchmarks/haarpy_benchmarks
    conda env create -f environment.haarpy_bench.yml
    conda activate haarpy_bench
    bash run_benchmarks.sh
    \end{lstlisting}
    The pinned conda environment specifies \texttt{haarpy==0.0.6}
    (overrideable at runtime via \texttt{HAARPY\_VERSION}).

    RTNI setup and run:
    \begin{lstlisting}[language=bash]
    # 1) Install RTNI for Mathematica and stage RTNI.wl + precomputedWG/
    #    in benchmarks/rtni_benchmarks/ (or on $Path)
    cd benchmarks/rtni_benchmarks
    export RTNI_EXPECTED_SHA256=cf7aaa1ba49b249ce7b9dd4a682bf2dd4cf83b47b8b92b341a84a882238694b2
    bash run_benchmarks.sh
    \end{lstlisting}
    For the manuscript comparison, the RTNI source revision is pinned by
    \texttt{RTNI\_EXPECTED\_SHA256} (full value shown in the command block
    above).
    The RTNI workflow records the package identifier in metadata (version symbol
    when available, otherwise \texttt{RTNI.wl} SHA-256).
    Each generated comparison result file includes an optional
    \texttt{\_meta} block (runtime/package versions, timestamp, host/OS
    details, and source identifiers) to document the exact run context across
    machines.
    \item \textbf{Manuscript build}: Build the paper from the \texttt{txt/}
    directory so relative figure/table paths resolve correctly:
    \begin{lstlisting}[language=bash]
    cd txt
    latexmk -pdf manuscript.tex
    \end{lstlisting}
\end{itemize}

\begin{acks}
ZP and {\L}P acknowledge support from the National Science Center (NCN), Poland,
under Project Opus No. 2022/47/B/ST6/02380.
\end{acks}

\bibliographystyle{ACM-Reference-Format}
\bibliography{references}

\end{document}